# Potential impacts of ballast water regulations on international trade, shipping patterns, and the global economy: An integrated transportation and economic modeling assessment

Zhaojun Wang[1], Duy Nong[2], Amanda M. Countryman[3,*], James J. Corbett[1], Travis Warziniack[4]

[1] *School of Marine Science and Policy, College of Earth, Ocean, and Environment, University of Delaware, Delaware, USA*
[2] *Agriculture and Food, The Commonwealth Scientific and Industrial Research Organisation, Australia*
[3] *Department of Agricultural and Resource Economics, Colorado State University, Colorado, USA*
[4] *USDA Forest Service Rocky Mountain Research Station, Colorado, USA*
**Corresponding author*
*Author e-mails: izhaojun@udel.edu (ZW), duy.nong@csiro.au (DN), Amanda.Countryman@colostate.edu (AMC), jcorbett@udel.edu (JJC), travis.w.warziniack@usda.gov (TW)*

## Abstract

Global ballast water management regulations aiming to decrease aquatic species invasion require actions that can increase shipping costs. We employ an integrated shipping cost and global economic modeling approach to investigate the impacts of ballast water regulations on bilateral trade, national economies, and shipping patterns. Given the potential need for more stringent regulation at regional hotspots of species invasions, this work considers two ballast water treatment policy scenarios: implementation of current international regulations, and a possible stricter regional regulation that targets ships traveling to and from the United States while other vessels continue to face current standards. We find that ballast water management compliance costs under both scenarios lead to modest negative impacts on international trade and national economies overall. However, stricter regulations applied to U.S. ports are expected to have large negative impacts on bilateral trade of several specific commodities for a few countries. Trade diversion causes decreased U.S. imports of some products, leading to minor economic welfare losses.

## 1. Introduction

Vessels routinely intake and discharge ballast water, corresponding to carried cargo, to maintain their stability. This activity transports organisms contained in ballast water from one place to another. If the discharge occurs outside the organisms' native range, there is potential for the species to become established and invasive. More than half of all marine invasive species are attributed to transfers by global commercial shipping (Molnar et al., 2008, Saebi et al., 2019), with around 10,000 species estimated to have been transported in ballast water (Bax et al., 2003). The impacts of invasive species on the environment and economic activities have been long investigated and found to impose negative impacts on the economy, ecosystem and human health (Carlton, 2003, Chan et al., 2019, Lovell et al., 2006, Wan et al., 2016, Wan et al., 2018, David et al., 2019).

Ballast water management (BWM) regulations attempt to reduce risk of species' spread and associated negative impacts. The International Convention on the Control and Management of Ship's Ballast Water and Sediments (the BWM Convention) of the International Maritime Organization (IMO), for example, defines allowable concentrations for viable organisms and certain human health indicator microbes contained in discharged ballast water in the D-2 Standard. While most countries are parties to the BWM Convention, the U.S. is not. Instead, the U.S. independently regulates ballast water discharge according to the 2018 Vessel Incidental Discharge Act (VIDA). Section 151.1511 or 151.2030 of Title 33, Code of Federal Regulations (or successor regulations) establishes the U.S. ballast water discharge standard, which is currently the same as the BWM Convention. However, individual states of the U.S. can set independent standards. California's interim performance standards, in effect from January 2030, establish limits for the number of organisms for different functional groups, which are the stricter than the BWM Convention[1]. California's final performance standard goal, in effect from January 2040, is to achieve zero detectable living organisms for all size classes. Current IMO standards are species concentration-based, instead of risk-based. Risks are not clearly mitigated by current standards, so there is a need for more stringent location-specific regulations at invasion hotspots (Verna and Harris, 2016, Saebi et al., 2019). To comply with the interim California performance standards, the California State Lands Commission funded a study to assess the feasibility of barge-based treatment methods (California State Lands Commission, 2018).

Compliance technology (i.e., type-approved ballast water treatment systems [BWTS]) are available to vessel operators to comply with both regulations and reduce species invasion risk. However, better systems may be necessary to achieve risk reduction targets. Procuring and installing BWTS on ships requires both capital and operating costs (King et al., 2009), and more advanced BWTS to comply with stricter regulations (if set) would cost more (Glosten et al., 2018). Vessel operators may reduce shipping services or costs may be passed on to cargo owners and eventually to final consumers (Tseng et al., 2005, Schinas and Stefanakos, 2012), which may negatively impact economic performance of particular sectors (Tseng et al., 2005). Herein, we assess the economic impacts of current international regulations and proposed location-specific alternatives including changes in international trade and transport (Estevadeordal et al., 2003, Hummels, 2007, Jacks et al., 2008). Shipping patterns serve as powerful tools in understanding environmental risk and analyzing policies, particularly those targeting marine bio-invasions and shipping emissions (Corbett et al., 2010, Wang et al., 2007, Johansson et al., 2017, Seebens et al., 2013). However, shipping patterns are dynamic, affected by economic policies (Halim et al., 2018a), climate change (Smith and Stephenson, 2013), and environmental regulations.

[1] California interim ballast water limits are as follows: no detectable living organisms greater than 50 micrometers in minimum dimension; less than 0.01 living organisms per milliliter less than 50 and more than 10 micrometers; less than 1000 bacteria and less than 10,000 virus per 100 milliliter, and different standards for Escherichia coli and Intestinal enterococci.

This work examines three aspects of impacts from current and future ballast water regulations: bilateral trade of specific commodities, overall impacts on national economies, and global shipping patterns by vessel type. This work can help inform risk-based assessments and allow comparisons of policies such as between current vessel-based treatment standards and additional measures to mitigate shipborne aquatic invasion risks. Our analysis informs evaluations of economic costs of decreasing risk of nonindigenous species spread in future policy contexts. Specific scenarios examined in this analysis are:

- Scenario 1 (Consistent BWM Regulation): This scenario is consistent with the current BWM regulation that all international and regional regulations require similar numeric standards of the BWM Convention.
- Scenario 2 (Stricter BWM Regulation): All ports in the U.S. adopt stricter BWM standards, following the regulation in California, while ports in all other countries apply the numeric standards of the BWM Convention.

We do not predict future policies but propose scenarios to explore the viability to drive policies forward. Choosing a U.S. based scenario adopting nationally the stringent standards of one state (California) is not without precedent. Many federal environmental policies since adoption of the Clean Water Act and Clean Air Act have essentially learned from and expanded standards initiated by one U.S. state. Moreover, our research uses the U.S. as an example in shipping policy, partly because there is precedent where international shipping adopted standards set by the U.S. One example is international double hull construction requirements for tank ships following U.S. enaction of the Oil Pollution Act of 1990. Lastly, insights from our analysis extend beyond investigation based on U.S. taking more stringent action (i.e., Singapore or Europe); this scenario design affords a major economy with hundreds of ports that trade with most of the major trading nations. We believe the study design is well suited to the economic-technology-environmental question of action to mitigate aquatic species invasion by ships. It is also important to note that our analysis does not evaluate or quantify the economic benefits of increased BWTS standards and corresponding decreased risk of nonindigenous species spread. The remainder of the paper is organized as follows. Section 2 highlights the methods and data. Section 3 provides the analysis of results and discussion, and Section 4 concludes this work.

## 2. Methodology

This research employs an innovative approach by integrating a shipping cost model and a global economic model to examine the impacts of ballast water management regulations on international trade, shipping traffic, and the overall economy. Figure 1 gives an outline of the data, model and outputs. The shipping cost model estimates changes in transport costs resulting from ballast regulation, which are employed as exogenous shocks to the marine transportation sector in the economic model to quantify changes in bilateral commodity trade and corresponding changes in national economies. Simulated changes in bilateral trade then serve as inputs for the shipping traffic accumulation model, which identifies potential changes in global shipping traffic. We include 20 countries[2] facing the highest compliance costs, most shipping voyages, and largest trade flows. We analyze nation-specific data and results for these countries and include all other countries as the aggregate *rest of the world* region.

---

[2] Australia, China, Japan, South Korea, Singapore, Malaysia, Taiwan, the United States, Canada, Mexico, Columbia, Panama, Venezuela, Belgium, Germany, Spain, France, the United Kingdom, the Netherlands, and South Africa.

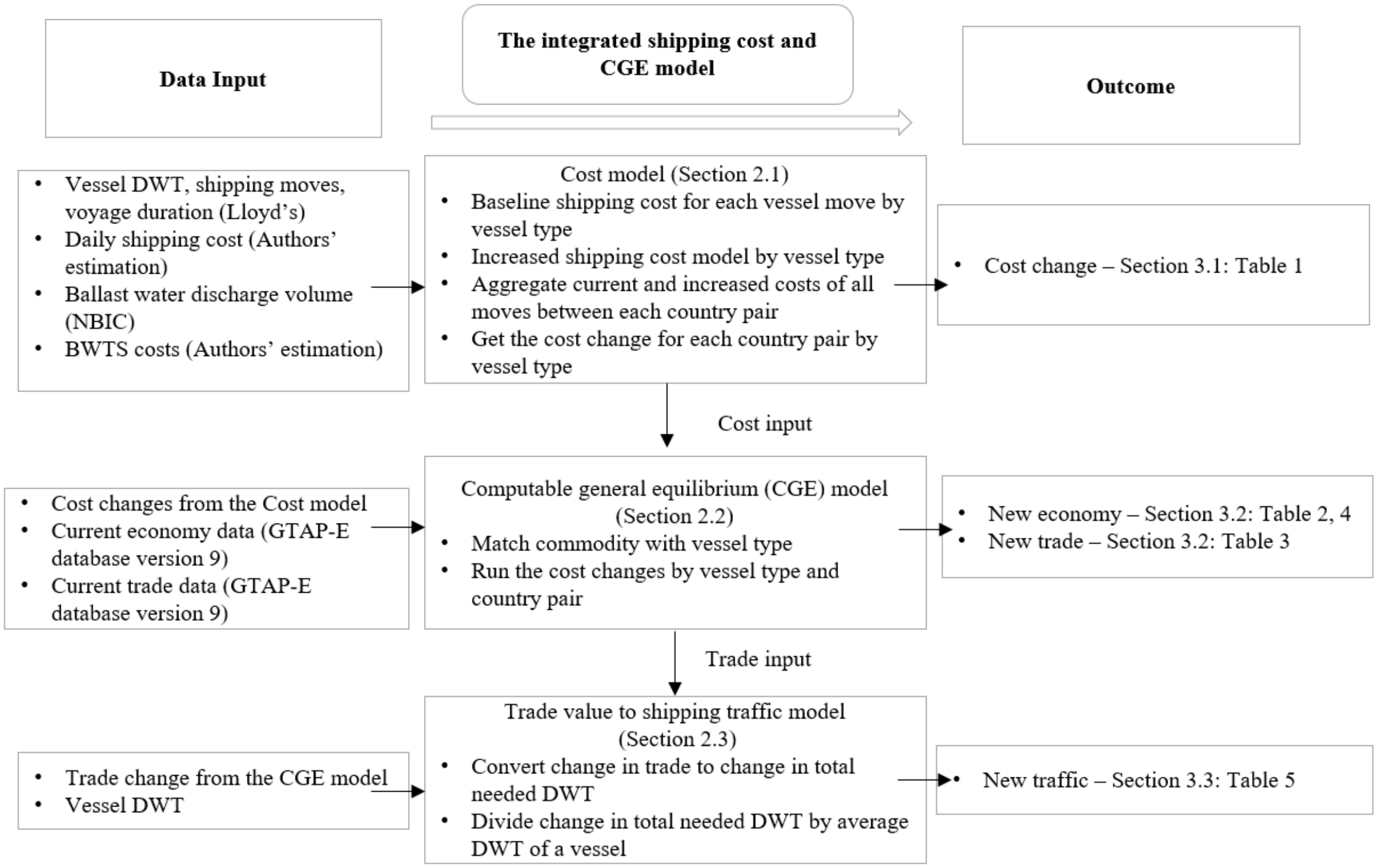

Figure 1. Outline of data, models, and outputs

## 2.1 Shipping cost model and data

The shipping cost model is composed of two parts: the current shipping cost model and the regulation compliance cost model. We calculate the shipping cost and compliance cost separately for each vessel type and every voyage. We then aggregate all voyages in one year to get the annual total current shipping cost and total compliance cost for each pair of countries. From this, we calculate the percentage increases in international shipping costs (the first component in Figure 1), which are used as exogenous shocks in the economic model.

### Baseline shipping cost for each vessel move

Shipping costs without ballast water management for each shipping voyage are the same over the two policy scenarios. The baseline shipping cost is comprised of capital, operating, and daily fuel costs. The cost components differ by vessel type; hence, the models are built separately for container, bulk, and tanker vessels. We estimate the total shipping costs for each vessel type with the daily shipping costs and voyage durations (See the descriptions in the Data section).

**Increased shipping cost due to regulatory compliance**

Baseline shipping costs increase across vessel types according to compliance costs of each regulation. In this regard, it is estimated that type-approved BWTS can be used to meet the regulatory standards required by the BWM Convention (labeled as IMO-BWTS). IMO-BWTS can be applied to vessels or barges. On the other hand, the barge-based BWTS assessed by the California State Lands Commission are intended to meet California's interim stricter standards (labeled as stricter-BWTS) as described in footnote 1 on page 2 (Glosten et al., 2018).

Because IMO-BWTS and barge-based BWTS generate different increases in shipping costs (i.e., the compliance costs), appropriate allocation of them to be vessel- or port-based can minimize the compliance cost. We identified the most economically efficient solutions for the world fleet in previous work (Wang and Corbett, 2020). That is, the fleet-wide most economically efficient pathway to comply with the Consistent Regulation scenario (Scenario 1) is to install vessel-based IMO-BWTS on every vessel; and the most economically efficient pathway to comply with the Stricter Regulation scenario (Scenario 2) is to use vessel-based IMO-BWTS at non-U.S. ports and use barge-based stricter-BWTS at U.S. ports. Then, with the following models, we compute the increased shipping costs due to regulatory compliance, which are composed of expenses associated with the BWTS capital, installation, operation, and ballast water treatment.

**(1) Regulatory compliance cost model under Consistent Regulation**

The compliance cost for each shipping voyage is obtained from Equation 1. Not all voyages calling at one port discharge ballast water, and we assume the discharge probability is 0.5 (Seebens et al., 2013). The capital and install costs are one-time costs used to calculate annual costs (lifetime of 30 years, a discount rate of 6%, and an annual inflation rate of 2.5%). Then we equally distribute annual costs to all voyages of that vessel because all the voyages use the onboard BWTS to treat discharged ballast water. The treatment cost for each voyage is obtained with the volume of discharged ballast water and unit treatment cost.

$$(C_{v-imo} + O_{v-imo})/N_v + T_{imo} * V_v \qquad \text{(Equation 1)}$$

Where

$C_{v-imo}$: annual capital cost and installation cost of vessel-based IMO-BWTS ($)

$O_{v-imo}$: annual operating cost for vessel-based IMO-BWTS ($)

$N_v$: the number of treatments of vessel $v$ in one year; the capital, installation, and operating costs are assumed to be shared by all treatments of that vessel

$T_{imo}$: treatment cost of IMO-BWTS for each treatment tonnage ($/ton)

$V_v$: ballast water treatment volume of vessel $v$ at that voyage (ton)

**(2) Regulation compliance cost model under Inconsistent Regulation**

Vessels often have different compliance costs because they call at different ports. We divide vessels into two groups: never call at U.S. ports, and may call at U.S. ports. Figure 2 shows the three kinds of voyages and corresponding equations used to calculate the compliance cost for each kind of voyage.

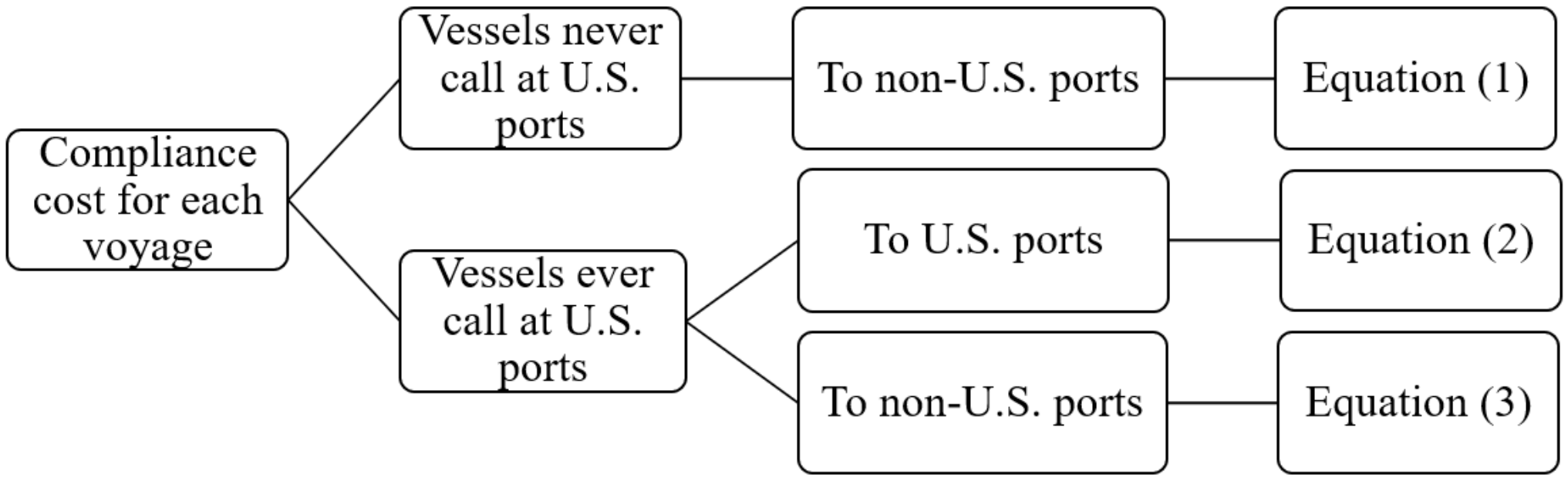

Figure 2. Three kinds of voyages determined by the vessel and voyage destination

If a vessel never calls at a U.S. port, it only needs to meet the IMO standard with the onboard IMO-BWTS, and the compliance cost is same as Equation 1.

If a vessel ever calls at a U.S. port, and the destination port of its voyage is a U.S. port, the discharged ballast water must meet stricter regulation, and the compliance cost calculation for that voyage follows Equation 2. We assume the costs of all barge-based BWTS are shared by voyages to U.S. ports, and the shared cost for each voyage is determined by the ballast water discharge volume of that voyage. The treatment cost is obtained according to the volume of discharged ballast water and the higher unit treatment cost. In addition, a cost component accounting for tug cost is added because barge-based BWTS need tugs to be moved and fixed.

$$(C_{barge} + C_{p-us} + O_{p-us}) * P_{us} * V_v / V_{all_US} + T_{us} * V_v + T_{tug} \quad \text{(Equation 2)}$$

Where

$C_{barge}$: annual capital cost and installation cost for one barge ($)

$C_{p-us}$: annual capital cost and installation cost for barge-based stricter-BWTS ($)

$O_{p-us}$: the sum of annual operating cost for one barge-based stricter-BWTS and one barge ($)

$P_{us}$: the number of ports in the U.S., which need stricter-BWTS at port

$V_{all_us}$: the total volume of treated ballast water at all U.S. ports per year (ton)

$T_{us}$: treatment cost of stricter-BWTS for each treatment tonnage ($/ton)

$T_{tug}$: the cost to use a tug per treatment ($/treatment)

If the vessel ever calls at a U.S. port, but the destination port of that voyage is not a U.S. port, the compliance cost estimation for that voyage follows Equation 3. For that voyage, the vessel must meet the IMO-standard with the onboard BWTS. Accordingly, that voyage shares a part of the vessel-based IMO-BWTS costs with the other voyages using it for treatment.

$$(C_{v-imo} + O_{v-imo})/N_{v_other} + T_{imo} * V_v \qquad \text{(Equation 3)}$$

Where

$N_{v_other}$: the number of treatments at non-U.S. ports of vessel *v* in one year.

To verify the compliance cost model, we sum compliance costs of all voyages to get the compliance cost for the world fleet. Then we compare that with the fleet cost obtained from the fleet compliance cost model (Wang and Corbett, 2020) and get the same number.

## Data

Data used for the cost model include shipping movements, daily shipping costs, ballast water discharge volume, and costs of BWTS. We use the data purchased from Lloyd's List Intelligence that gives movement of vessels throughout the world prior to potential treatment cost. The data include Vessel ID, Origin/Destination port, Port ID, Origin/Destination country, Departure/Arrival time, along with vessel specifications (Vessel Type, Deadweight Tonnage (DWT), Year of build, and other descriptive data). Since the BWM Convention applies only to ships with international voyages, we exclude movements of ships that travel within a country. The final dataset includes 714,039 individual vessel moves.

The voyage durations are obtained from Lloyd's List Intelligence. Daily shipping costs are derived from the Guide to Deep-Draft Vessel Operating with a 2002 base year (US Army Corps of Engineers, 2002), following the literature (Corbett et al., 2009, Dumas and Whitehead, 2008, The EPA, 2012). To verify that these daily shipping costs reflect current costs, we run the recently established container vessel shipping cost model for vessels in sizes of 600, 1600, 2500, 4000, and 6000 TEU and get similar daily costs (Wang et al., 2020). We also confirm that capital costs derived from the USACE publication are consistent with current costs using long-run analyses by the OECD (OECD and BRS Group, 2018). We believe this demonstrates that overcapacity in shipbuilding continues to impact market cycles and demand pricing, and partly because unit costs often decline over time for technology. If higher shipping cost baselines were used in the analysis, the relative effect of BTWS costs would be smaller. With these baseline shipping costs, we believe the relative effect of added ballast water treatment costs provides reasonable and conservative estimates for exogenous shocks to the global economic model.

We estimate ballast water discharge volume for each shipping voyage with a regression following Seebens et al. (2013), with ballast water discharge data published by the National Ballast Information Clearinghouse (National Ballast Information Clearinghouse, 2019, Seebens et al., 2013). Not all ships discharge ballast water when entering a port; the fraction of port arrivals without ballast water discharge is estimated to be between 42% and 88%, depending on vessel type (Seebens et al., 2013). Hence, we assume a ballast water discharge probability of 50% in this study. For vessels with missing DWT information, we use an average discharge volume to estimate ballast discharge volumes.

Several works estimate the cost of barge-based BWTS (Glosten et al., 2018, COWI A/S, 2012, Maglić et al., 2015, King and Hagan, 2013), and we use cost estimates from Glosten et al. for California to achieve stricter-standard (Glosten et al., 2018). Though a barge-based BWTS industry is not in place, the Glosten et al. (2018) report is the best estimate that can be obtained. The work of King and Hagan provides cost data for the IMO-standard BWTS (King et al., 2009). These two reports include various BWTS with different cost estimates for different vessels. This work does not aim to identify the best BWTS for each vessel; rather, we investigate costs of the whole world fleet. Therefore, we use the average of the highest and lowest values from the literature to represent the best estimates. The lower cost values are within 0.75 times of the average and higher cost values are within 1.5 times of the average.

## 2.2 Economic model and data

### Global Trade Analysis Project (GTAP) model

This work uses a computable general equilibrium (CGE) modeling framework known as GTAP (Global Trade and Analysis Project), specifically an extension of the GTAP-Energy/Environment (GTAP-E) model by Nong and Siriwardana (2017). The GTAP-E model is widely used and has been employed to examine the impacts of various policies on international trade (for example, see Baier and Bergstrand (2009); Bekkers et al. (2016); Countryman et al. (2016)). CGE models include interactions between producers, consumers, investors, households, and governments. Though CGE models require more calibration data and behavioral assumptions than sector-specific economic models, CGE models are useful when relationships between sectors or countries are important or when policies imposed on one sector have economy-wide impacts, which are all likely to be true when considering policies targeting transportation and shipping sectors. A brief description of the model and specific changes made for this analysis area given here; more details on the model structure can be found in Nong and Siriwardana (2017).

Economic agents in the model include households, government, industries, and investors. Households and governments are the final consumers in each country; they buy goods and services supplied by domestic and international industries. Households maximize utility subject to their budget constraints. A household receives income by supplying labor, and the government receives tax revenue (sales tax, business taxes, import and export taxes, and investment taxes). Industries provide outputs for final consumers and goods used in the production process of other sectors. The production process in each sector uses goods produced domestically and from international markets, labor, capital, and other resources. In addition, national economies are connected by bilateral trade for each commodity in the model.

The model includes transportation services, such as water, air, road, and rail, which are used to facilitate trade between sectors and countries. Industries use transportation services to move products to other sectors in domestic and international markets. For example, agricultural industries in China may use road and rail transport services to transport agricultural goods to customers in India, while air and/or water transport services are used to move products to customers in the U.S. or Canada. Agricultural industries in China also use transportation services to move products to other industries and final customers in their domestic market.

In this model, changes in transport costs affect consumer and industry purchasing power, thereby affecting trade volumes between sectors and countries. That is, demands for commodities will change when commodity prices change. For example, if transportation costs from China to the U.S. fall, and other conditions held constant, U.S. customers will substitute away from relatively higher priced suppliers and purchase cheaper products from China. All agents are connected; therefore, changes in transport costs associated with ballast water treatment lead to changes in global trade patterns.

We use the GTAP-E database version 9, which represents the world economy in 2011 (Aguiar et al., 2016) in tandem with the modified GTAP-E model. The global database includes production, consumption, investment, import, and export values[3]. We aggregate 134 regions into 21 regions as discussed in Section 2.1 and aggregate 57 sectors/commodities into 17 main sector/commodity groups, which are mapped with 8 vessel types as described in Appendix A.

### 2.3 Trade to shipping traffic

The next step is to estimate changes in shipping traffic for container, bulk, and tanker vessels given the results of the economic model (the lowest part in Figure 1). Accordingly, we estimate changes in shipping voyages for different vessel types with the DWT of vessels engaged in trade and allocated to a route (Equation 4), which are determined by the simulated changes in total trade from the economic model (Equation 5). The total DWTs in the equations represent demand for total DWT of the working world fleet of each vessel type.

$$\text{Change in shipping voyages} = \frac{\text{change in total allocated DWT}}{\text{vessels' average DWT}} \qquad \text{(Equation 4)}$$

$$\text{Value/weight ratio} = \frac{\text{current total trade value}}{\text{current total cargo volume}} = \frac{\text{change in trade value}}{\text{change in cargo volume}}$$

[3] The database is measured in U.S. dollars. Exchange rates do not play any role in this modeling.

$$= \frac{\text{current total trade value}}{\text{current allocated DWT} * \alpha} = \frac{\text{change in trade value}}{\text{change in total allocated DWT} * \alpha} \quad \text{(Equation 5)}$$

First, value/weight ratios are estimated for different commodities between different countries (Halim et al., 2018a, Halim et al., 2018b) to serve as a bridge to link the status quo and changes due to regulatory compliance for the same commodity between the same country pair. Second, we assume the utilization of a shipping route is constant, i.e., the cargo volume carried by a vessel is assumed to be constant with the proportion of its DWT.

## 3. Results and discussion

We present findings for changes in shipping costs, followed by the economic impacts on trade and national economies, and changes in global shipping traffic as a result of BWM regulations defined by Scenarios 1 and 2. Transportation industries account for small contributions to the global economy in terms of monetary values, which leads to modest aggregate economic effects. However, we find substantial impacts on certain commodities, countries, and shipping routes.

### 3.1 Changes in shipping costs

Changes in shipping costs for container vessels, bulk vessels, and tankers are provided in Tables S1–S6 in the Supplementary Material. The average compliance cost (per vessel) of these three vessels accounts for 1.5% of the current shipping cost under Scenario 1, and 2% under Scenario 2. These percentages become 4.6% and 4.9% when all vessel types are included (Wang and Corbett, 2020). The current shipping cost is composed of operating cost (14%), periodic maintenance (4%), voyage cost (40%), and capital cost (42%) (Stopford, 2009). The average ballast water management compliance cost is lower than the periodic maintenance cost.

Shipping costs for voyages between most countries in both scenarios do not change substantially; however, costs for routes between some country pairs change greatly. Therefore, we report the pairs of countries with large cost changes, i.e., larger than 10% for at least one vessel type, in Table 1. Overall, Scenario 2, which provides better protection from species invasion, leads to the largest increases in shipping costs across vessel types.

Table 1. Changes in shipping costs in 2011 for three vessel types traveling between countries

| Pair of countries | Container: Scenario 1 | | Container: Scenario 2 | | Bulk: Scenario 1 | | Bulk: Scenario 2 | | Tanker: Scenario 1 | | Tanker: Scenario 2 | |
|---|---|---|---|---|---|---|---|---|---|---|---|---|
| | % | $ Thousand | % | $ Thousand | % | $ Thousand | % | $ Thousand | % | $ Thousand | % | $ Thousand |
| CAN/USA | 2 | 800 | 15 | 7,800 | 3 | 4,000 | 36 | 46,000 | 2 | 1,600 | 14 | 16,000 |
| COL/USA | 1 | 150 | 8 | 1,300 | 2 | 560 | 15 | 5,100 | 1 | 890 | 9 | 8,900 |
| MEX/USA | 1 | 750 | 11 | 7,700 | 3 | 2,200 | 27 | 21,000 | 2 | 3,600 | 16 | 39,000 |
| PAN/USA | 1 | 1,800 | 6 | 13,000 | 2 | 3,100 | 16 | 25,000 | 1 | 2,300 | 8 | 19,000 |
| VEN/USA | 1 | 73 | 11 | 750 | 2 | 500 | 11 | 500 | 1 | 1,800 | 8 | 19,000 |
| ESP/MEX | 11 | 5 | 11 | 5 | 1 | 3 | 1 | 3 | 1 | 10 | 1 | 11 |
| MEX/VEN | 23 | 3 | 23 | 3 | 1 | 22 | 5 | 27 | 1 | 34 | 2 | 41 |
| AUS/USA | 7 | 2 | 53 | 11 | 1 | 210 | 4 | 1250 | 1 | 18 | 2 | 133 |
| BEL/AUS | - | - | - | - | 22 | 3 | 23 | 4 | - | - | - | - |
| VEN/AUS | - | - | - | - | 25 | 5 | 25 | 5 | - | - | - | - |
| BEL/GBR | - | - | - | - | 10 | 1,200 | 11 | 1,300 | 4 | 1,990 | 4 | 2080 |
| DEU/VEN | - | - | - | - | 19 | 2 | 19 | 2 | - | - | - | - |
| NLD/VEN | - | - | - | - | 23 | 3 | 24 | 3 | - | - | - | - |
| DEU/USA | 1 | 50 | 3 | 520 | 1 | 65 | 9 | 720 | 1 | 8 | 14 | 76 |
| PAN/MYS | - | - | - | - | 1 | 4 | 1 | 5 | 23 | 2 | 24 | 2 |
| CHN/PAN | 1 | 540 | 1 | 680 | 1 | 1100 | 1 | 1230 | 22 | 4 | 23 | 4 |
| ESP/SGP | - | - | - | - | 1 | 7 | 1 | 7 | 14 | 300 | 14 | 300 |
| KOR/VEN | - | - | - | - | - | - | - | - | 15 | 2 | 15 | 2 |
| ZAF/VEN | - | - | - | - | 1 | 9 | 1 | 9 | 15 | 5 | 16 | 5 |

Note: Scenario 1 corresponds to Consistent IMO BWM Regulation and Scenario 2 represents the U.S. Stricter BWM Regulation. Percentages rounded to integral for reader comparison; percentages less than 1% are round to 1%. Dollars rounded to thousand. The routes shown have at least a 10% change in cost for one of the ship types. No traffic may between the routes for certain vessel types, shown with -. Countries are shown in three-digits Code: https://unstats.un.org/unsd/tradekb/knowledgebase/country-code.
Source: Authors' calculations

The high percentage changes are a result of high compliance costs or relatively low baseline shipping costs. For example, the percentage change in container vessel shipping from Belgium to Australia in Table 1 is as high as 22%, but its absolute cost change is only $3,000. In contrast, the percentage change for a container vessel voyage from Belgium to the UK is less than half of the aforementioned value (10%), but the absolute cost change is $1.2 million. This is because of the relatively high baseline shipping cost from Belgium to the UK and lower baseline shipping cost from Belgium to Australia. The baseline shipping cost from Belgium to the UK is higher than that of Belgium to Australia, even though the distance is shorter, because there are a greater number of voyages that occur. The country-to-country shipping cost is aggregated over all shipping voyages between a country pair.

Table 1 indicates more country pairs have large changes in shipping costs for bulk vessels and tankers. The average change in shipping costs of tankers are 3.4% and 3.9% under Scenario 1 and 2, respectively. The changes are 1.8% and 2.4% for bulk vessels, and 1.3% and 1.7% for container vessels. This reveals that ballast water discharge regulations fall more heavily on bulk vessels and tankers, and this is true for both Consistent IMO Regulation (Scenario 1) and Stricter Regulation (Scenario 2). This is attributed to several possible reasons: First, bulk vessels and tankers have fewer annual voyages per vessel on average (9 and 13 voyages), compared to container vessels (18 voyages), so the cost of vessel-based BWTS is shared among fewer voyages, causing higher cost changes[4]. Second, bulk vessels and tankers discharge more ballast water in total (389 and 280 million tons) compared to container vessels (108 million tons), so bulk vessels and tankers have higher treatment costs under both scenarios.

Changes in shipping costs under the two scenarios are similar in the magnitude for country pairs other than the U.S., indicating that regionally stricter ballast water regulation and better protection from aquatic invasion has minimal impact on non-U.S. routes. Relatively small differences that arise for some country pairs can be explained with the cost models. Taking a country pair, A-B for example, the shipping cost calculation only includes the voyages from A to B of all vessels. However, some vessels undertake many voyages and travel to other countries besides B. If all the destinations are non-U.S. countries, then the costs are identical in Scenarios 1 and 2. This is because the vessel only needs to meet the IMO standard with its onboard BWTS, and the increased compliance cost refers to the cost of the BWTS that does not change across the two regulatory scenarios. If the vessel travels to U.S. ports besides B, then shipping costs increase more under Scenario 2. This is because under Scenario 2, the vessel must use port-based BWTS to treat ballast water at U.S. ports to achieve the stricter standards and use vessel-based BWTS at other ports. In this way, the compliance cost includes some of the port-based cost and its onboard BWTS. However, under Scenario 1, the compliance cost only refers to the cost of its vessel-based BWTS, which is lower than costs incurred under Scenario 2.

[4] The data are from authors' calculations.

### 3.2 Impacts on International Trade and National Economies

The impacts of increased costs from BWM in Scenarios 1 and 2 cause relatively minor changes to aggregate trade levels across countries; however, noteworthy decreases in bilateral trade of certain commodities between trade partners occurs. In Scenario 1, when all countries follow the international standard to the BWM Convention, the induced costs are relatively small, as fully described previously, leading to corresponding minor impacts on overall trade between all countries. Columns (1) and (5) in Table 2 show that imports and exports for most countries decline by less than 0.1%. Stricter standards outlined in Scenario 2 lead to larger changes in trade than Scenario 1, and trade between the U.S. and its primary trading partners are affected the most. Columns (3) and (4) in Table 2 show that exports from Australia, the U.S., and Canada decline by 0.14% ($380 million), 0.31% ($5.7 billion), and 0.25% ($1.2 billion), respectively. In addition, imports into several countries also decrease by 0.27% ($700 million) for Australia, 0.33% ($8.8 billion) for the U.S., 0.4% ($1.9 billion) for Canada, 0.24% ($770 million) for Mexico, and 0.34% ($180 million) for Venezuela.

Although aggregate, country-level, changes in trade are minor, there are noteworthy disruptions in bilateral trade of several commodities resulting from increased BWM regulatory compliance costs. Table 3 provides changes in bilateral commodity trade shipped by container, bulk, and tanker vessels. In general, changes in bilateral exports of commodities are highly and negatively correlated with changes in shipping costs of vessels, as expected. For example, in Scenario 1, the shipping cost of container vessels from China to Venezuela increases by 8.83% (Table S1 in Supplementary Material), and exports of commodities transported by container vessels for this route decrease by -1 to -2.6% (Table 3). By contrast, shipping costs of container vessels from Mexico to China only increase by 0.13% in Scenario 1 (Table S1), leading to a 0.1% change in exports transported by container vessels for this route (Table 3).

Table 2. Changes in real aggregate imports and exports by country

| Countries | Exports | | | | Imports | | | |
|---|---|---|---|---|---|---|---|---|
| | Scenario 1 | | Scenario 2 | | Scenario 1 | | Scenario 2 | |
| | % change (1) | Real change ($ million) (2) | % change (3) | Real change ($ million) (4) | % change (5) | Real change ($ million) (6) | % change (7) | Real change ($ million) (8) |
| Australia | -0.04 | -110 | -0.14 | -380 | -0.08 | -220 | -0.27 | -700 |
| China | -0.06 | -1,200 | -0.05 | -950 | -0.1 | -1,700 | -0.08 | -1,300 |
| Japan | -0.03 | -300 | -0.01 | -130 | -0.05 | -510 | -0.02 | -210 |
| South Korea | -0.02 | -150 | 0 | 8 | -0.04 | -250 | -0.01 | -42 |
| Singapore | -0.02 | -59 | 0 | -6 | -0.02 | -45 | 0.02 | 58 |
| Malaysia | -0.03 | -80 | -0.02 | -42 | -0.07 | -140 | -0.05 | -100 |
| Taiwan | -0.03 | -120 | -0.02 | -90 | -0.06 | -170 | -0.04 | -120 |
| USA | -0.03 | -580 | -0.31 | -5,700 | -0.04 | -990 | -0.33 | -8,800 |
| Canada | -0.05 | -250 | -0.25 | -1,200 | -0.08 | -400 | -0.4 | -1,900 |
| Mexico | -0.02 | -88 | -0.06 | -200 | -0.07 | -220 | -0.24 | -770 |
| Colombia | -0.02 | -10 | -0.04 | -20 | -0.06 | -34 | -0.19 | -110 |
| Panama | -0.05 | -9 | -0.13 | -22 | -0.05 | -18 | -0.11 | -37 |
| Venezuela | -0.03 | -25 | -0.06 | -45 | -0.14 | -71 | -0.34 | -180 |
| Belgium | 0.01 | 43 | 0.03 | 140 | 0.01 | 47 | 0.04 | 160 |
| Germany | 0 | -30 | 0.01 | 160 | 0 | 6 | 0.02 | 320 |
| Spain | -0.01 | -23 | 0 | 7 | -0.01 | -49 | 0 | -8 |
| France | 0.01 | 81 | 0.03 | 210 | 0.01 | 76 | 0.03 | 220 |
| UK | -0.04 | -290 | -0.01 | -80 | -0.05 | -420 | -0.02 | -170 |
| Netherlands | 0 | -17 | 0.02 | 74 | -0.01 | -39 | 0.02 | 65 |
| South Africa | -0.01 | -15 | 0 | -5 | -0.03 | -31 | -0.01 | -16 |
| Rest of World | -0.03 | -2,400 | -0.04 | -2,900 | -0.06 | -4,300 | -0.08 | -5,700 |
| The whole world | -0.03 | -5,600 | -0.06 | -11,000 | -0.05 | -9,500 | -0.1 | -19,000 |

Note: Scenario 1 corresponds to the Consistent IMO BWM Regulation and Scenario 2 represents the U.S. Stricter BWM Regulation. Export values are measured at FOB (free on board) prices, which include the costs of delivering goods to the ports. Import values are measured in CIF (cost insurance and freight) prices, which include transportation costs and insurance fees to having goods to the port of destinations. Percentages rounded to two decimals, and dollars rounded to millions, for reader comparison; smaller numbers do not imply precision.
Source: Authors' simulations

Table 3: Changes in exports by commodity between countries.

| Vessel types | Commodities | Australia to USA | | | | China to Venezuela | | | | USA to Canada | | | | USA to Mexico | | | |
|---|---|---|---|---|---|---|---|---|---|---|---|---|---|---|---|---|---|
| | | Scenario 1 | | Scenario 2 | | Scenario 1 | | Scenario 2 | | Scenario 1 | | Scenario 2 | | Scenario 1 | | Scenario 2 | |
| | | (a) | (b) | (a) | (b) | (a) | (b) | (a) | (b) | (a) | (b) | (a) | (b) | (a) | (b) | (a) | (b) |
| Bulk | Crop products | 0.2 | 0 | 2.7 | 0.3 | 0 | 0 | -0.4 | 0 | -1 | -3.4 | -2.9 | -9.4 | -0.2 | -0.5 | -1.7 | -4.9 |
| Bulk | Wheat and grains | 0.3 | 0 | 5.7 | 0 | 0.4 | 0 | 0.1 | 0 | -0.8 | -3.1 | -1.8 | -6.8 | -0.3 | -14 | -0.8 | -32.3 |
| Bulk | Coal | 0.4 | 0 | 5.3 | 0.2 | 0.3 | 0 | -0.2 | 0 | -1.8 | -8.1 | -3.7 | -16.6 | -1.4 | -2.4 | -2.3 | -4.1 |
| Container | Forestry and foods | -1.6 | -16.5 | -21.1 | -220.9 | -2.3 | -0.4 | -2.5 | -0.4 | -0.2 | -21.1 | -0.6 | -63.8 | -0.3 | -19.7 | -0.6 | -42.9 |
| Container | Textiles | -2.7 | -6.7 | -33.9 | -83.4 | -2.6 | -22.1 | -2.9 | -24.7 | -0.3 | -44.3 | -1 | -142.4 | -0.2 | -18.1 | -0.5 | -56.2 |
| Container | Metal & chemicals | -1.3 | -46.1 | -18.8 | -686.9 | -2.6 | -27.3 | -2.8 | -29.5 | -0.3 | -160.3 | -0.8 | -502.6 | -0.1 | -47.9 | -0.4 | -207.8 |
| Container | Machine & equip. | -0.7 | -21.5 | -12.1 | -368.2 | -1 | -36.9 | -1.1 | -42.6 | -0.1 | -57.2 | -0.5 | -284.6 | 0 | -12.7 | -0.3 | -119.1 |
| Tanker | Oil products | 0.1 | 0.1 | 0.9 | 0.8 | -0.1 | 0 | -0.2 | 0 | 0.2 | 22.8 | -0.5 | -59.5 | -0.2 | -47 | -0.8 | -164 |
| Tanker | Gas | 1.3 | 0.2 | 3.9 | 0.5 | 0.5 | 0 | -0.6 | 0 | -0.5 | -12.1 | -1.4 | -31.1 | -0.7 | -12.8 | -2.3 | -44.2 |
| Tanker | Crude oil | 0.2 | 0.4 | 1.1 | 2.4 | 0.4 | 0 | -0.2 | 0 | -0.5 | -1.8 | -2.4 | -8 | -3 | 0 | -6.5 | 0 |

| Vessel types | Commodities | Canada to USA | | | | Spain to Mexico | | | | Colombia to China | | | | Colombia to USA | | | |
|---|---|---|---|---|---|---|---|---|---|---|---|---|---|---|---|---|---|
| Bulk | Crop products | 0 | 0.1 | -1.6 | -7.4 | 0.2 | 0 | -0.3 | 0 | 0.1 | 0 | 1.9 | 0 | -0.7 | -12.8 | -6.3 | -119.3 |
| Bulk | Wheat and grains | -0.4 | -5.2 | -6.2 | -82.5 | 0.8 | 0 | 1 | 0 | 0.1 | 0 | 0.6 | 0 | -0.1 | 0 | 2.3 | 0 |
| Bulk | Coal | -0.8 | -1.5 | -14.4 | -28.1 | 0.5 | 0 | 0.4 | 0 | 0.3 | 0.5 | 0.8 | 1.3 | -0.3 | -2.1 | -2 | -16.4 |
| Container | Forestry and foods | 0 | -1 | 0.2 | 17.8 | -2 | -4.2 | -2.1 | -4.4 | 0.1 | 0 | 0.4 | 0 | -0.2 | -0.9 | -1 | -5.4 |
| Container | Textiles | -0.3 | -60.4 | -3 | -662.8 | -2.7 | -17.9 | -2.7 | -17.5 | 0.1 | 0 | 0.6 | 0.2 | -0.1 | -0.6 | -0.8 | -3.3 |
| Container | Metal & chemicals | -0.2 | - [illegible] | -1.8 | -1220 | -2.8 | -29.4 | -2.8 | -29.1 | 0.3 | 1.7 | 0.7 | 4.1 | 0 | -1.2 | 0 | 0.1 |
| Container | Machine & equip. | -0.1 | -34.3 | -0.7 | -251.7 | -1 | -13.9 | -0.9 | -13.1 | 0 | 0 | 0.6 | 0 | -0.1 | -0.3 | -0.3 | -1 |
| Tanker | Oil products | -0.3 | -31.7 | -2.5 | -316.3 | 0.2 | 0.1 | 0.5 | 0.3 | 0.2 | 0 | 0.7 | 0 | -0.1 | -0.5 | -0.8 | -8 |
| Tanker | Gas | 0.1 | 7.7 | 1.6 | 180.8 | 2 | 0 | 2.8 | 0 | 0.1 | 0 | 3.2 | 0 | 0.4 | 0 | 5.5 | 0 |
| Tanker | Crude oil | 0 | -19.6 | 0 | 4.7 | -0.2 | 0 | -1.7 | 0 | 0.2 | 1.3 | 1.3 | 10.8 | -0.1 | -15.2 | -1 | -103.6 |

| Vessel types | Commodities | Mexico to | | | | | | | | | | | | | | | |
|---|---|---|---|---|---|---|---|---|---|---|---|---|---|---|---|---|---|
| | | China | | | | USA | | | | Venezuela | | | | Spain | | | |
| Bulk | Crop products | -0.3 | 0 | 1.4 | 0 | -0.6 | -3.2 | -7 | -38.5 | -1.3 | 0 | -0.3 | 0 | 0.1 | 0 | 1.8 | 0 |
| Bulk | Wheat and grains | 0.1 | 0 | 0.8 | 0 | -0.5 | -0.1 | -3.4 | -0.6 | -0.7 | -0.5 | -0.5 | -0.4 | 0 | 0 | 0.7 | 0 |
| Bulk | Coal | -0.9 | 0 | -0.5 | 0 | 0.6 | 0 | 7.8 | 0 | 0.3 | 0 | 0.6 | 0 | 0.2 | 0 | 0.8 | 0 |
| Container | Forestry and foods | 0.1 | 0.2 | 0.5 | 0.9 | -0.1 | -8.6 | -0.8 | -67.6 | -7.6 | -4.9 | -7.5 | -4.8 | 0 | 0.5 | 0.3 | 0 |
| Container | Textiles | 0.1 | 0.2 | 0.6 | 1.1 | 0 | -2.3 | 0.1 | 16.3 | -10.9 | -7.6 | -10.7 | -7.5 | 0 | 0.2 | 0.4 | 0 |
| Container | Metal & chemicals | 0.1 | 4 | 0.5 | 16.7 | -0.1 | -17.5 | -0.7 | -213.5 | -4.4 | -40 | -4.1 | -37.6 | 0.1 | 1.6 | 0.5 | 0.2 |
| Container | Machine & equip. | 0.1 | 1.4 | 0.7 | 11 | 0 | 15.4 | 0.1 | 95.7 | -3.1 | -13.8 | -2.7 | -12 | 0.1 | 1.8 | 0.6 | 0.1 |
| Tanker | Oil products | 0.2 | 0 | 0.9 | 0.1 | -0.2 | -9.9 | -2 | -113.7 | 0 | 0 | 0.5 | 0 | 0.1 | 0 | 0.9 | 0 |
| Tanker | Gas | -0.5 | 0 | 3.4 | 0 | -0.5 | -0.1 | 1.3 | 0.1 | 0.1 | 0 | 3.4 | 0 | -0.5 | 0 | 3.5 | 0 |
| Tanker | Crude oil | 0 | 0.1 | 3.1 | 26.7 | -0.3 | -65.2 | -2.3 | -614.5 | 0.6 | 0 | 3.4 | 0 | 0.3 | 90.6 | 3.2 | 7.2 |

| Vessel types | Commodities | Venezuela to | | | | | | | | | | | | | | | |
|---|---|---|---|---|---|---|---|---|---|---|---|---|---|---|---|---|---|
| | | China | | | | USA | | | | Belgium | | | | Germany | | | |
| Bulk | Crop products | 0.1 | 0 | 0.8 | 0 | -0.4 | 0 | -0.3 | 0 | -0.2 | 0 | 0.2 | 0 | 0 | 0 | 0.5 | 0 |
| Bulk | Wheat and grains | 0.1 | 0 | 0.6 | 0 | 0.2 | 0 | 6.4 | 0 | 0 | 0 | 0.4 | 0 | -0.1 | 0 | 0.4 | 0 |
| Bulk | Coal | 0.3 | 0 | 0.7 | 0 | -0.6 | -0.1 | -0.8 | -0.2 | -0.1 | 0 | 0.3 | 0 | -0.4 | -0.1 | -0.1 | 0 |
| Container | Forestry and foods | 0 | 0 | 0.4 | 0 | -0.2 | -0.1 | -1.3 | -0.6 | 0 | 0 | 0.2 | 0 | -0.1 | 0 | 0.2 | 0 |
| Container | Textiles | -0.1 | 0 | 0.5 | 0.1 | -0.1 | 0 | 1.2 | 0.1 | -0.2 | 0 | 0.4 | 0 | -0.2 | 0 | 0.3 | 0 |
| Container | Metal & chemicals | 0 | -0.3 | 0.4 | 4.9 | -0.7 | -7.9 | -5.4 | -66.1 | -0.1 | -0.3 | 0.4 | 1.2 | -0.1 | -0.2 | 0.4 | 0.9 |
| Container | Machine & equip. | -0.2 | 0 | 0.6 | 0 | -0.5 | -0.4 | -3.2 | -2.2 | -0.2 | 0 | 0.5 | 0 | -0.2 | 0 | 0.5 | 0 |
| Tanker | Oil products | 0.1 | 7.1 | 0.5 | 26.7 | -0.1 | -2.8 | -0.6 | -34.2 | 0.1 | 0.1 | 0.6 | 0.4 | 0 | 0 | 0.5 | 0.5 |
| Tanker | Gas | -0.5 | 0 | 1.4 | 0 | -0.2 | 0 | 3.7 | 0 | -1.1 | 0 | 1.1 | 0 | -0.5 | 0 | 1.4 | 0 |
| Tanker | Crude oil | 0 | 1.9 | 1 | 52.8 | -0.2 | -43.6 | -1 | -272.4 | 0.2 | 2.5 | 1.3 | 13.1 | 0.3 | 3 | 1.3 | 12.6 |

Note: (a) indicates percentage change, (b) indicates value change ($ million). Numbers rounded to one decimal. Scenario 1 corresponds to the Consistent IMO BWM Regulation and Scenario 2 represents the U.S. Stricter BWM Regulation. Results include both absolute value changes greater than $10 million and percentage changes greater than 1% for at least one vessel type.

Source: Authors' simulations

Results for Scenario 1 in Table 3 indicate that if all international and regional regulations require similar standards to the BWM Convention, there will be negligible impacts on most bilateral trade in the international market. For example, exports of forestry and food products, and metal and chemicals commodities from Australia to the U.S. only decrease by 1.6% ($16.5 million) and 1.3% ($46.1 million), respectively. In Scenario 2, when all ports in the U.S. apply stricter BWM regulation, economic impacts are higher and spread across more commodities and countries (Table 3). For example, exports of textiles, metal and chemicals, and machines and equipment from China to Venezuela decline by 2.6% ($22.1 million), 2.6% ($27.3 million), and 1% ($36.9 million) in Scenario 1, respectively. These exports fall by 2.9% ($24.7 million), 2.8% ($29.5 million), and 1.1% ($42.6 million) in Scenario 2, respectively. This is because higher transportation costs between the U.S. and all other countries in Scenario 2 cause trade diversion resulting from changes in relative shipping costs. Exports from some countries to the U.S. decrease substantially in Scenario 2. Specifically, exports from Australia to the U.S. decline at relatively high rates. For example, exports of the other agricultural commodities product category decrease by 21.1% ($220.9 million), textiles by 33.9% ($83.4 million), metal and chemicals by 18.8% ($686.9 million), and machine and equipment by 12.1% ($368.2 million). Trade diversion is illustrated when exports of metal and chemicals from Australia to the U.S. decrease, as previously described, yet increase to China (+0.14% or +$80 million), Japan (+0.37% or +$62 million), and South Korea (+0.5% or +$62 million) (Table S7 in Supplementary Material). In this regard, stricter regulation applied to all U.S. ports adversely affects exports to the U.S., as expected. Trade diversion also affects domestic and foreign production. For example, the impacts on Australian sectors' performance are relatively higher in Scenario 2 (Table S8 in Supplementary Material).

There are mixes of both positive and negative changes because of the substitution effects when global transport costs increase. That is, countries will divert purchases from higher cost suppliers with lower cost sources. Also, there are sectoral level substitution effects in each country where low cost inputs are used in place of relatively higher cost commodity inputs. For example, sectors can substitute between coal, petroleum products, natural gas, and crude oil when prices of any commodities increase relative to the others. Table S8 also shows that the production levels of sectors in the U.S. are more negatively affected in Scenario 2. This is because U.S. sectors face higher costs for imported commodities, thereby increasing production costs. Some sectors, however, experience additional gains because industrial, private and public sectors substitute lower cost inputs to replace higher cost commodity inputs. For example, the crop product and crude oil sectors increase output levels by 0.51% and 0.17% in Scenario 2, respectively, compared to 0.03% and 0.01% in Scenario 1. In this instance, crude oil can be substitutable for coal, petroleum products, and natural gas, while crop products are used as substitutes for other agricultural products.

Results indicate macroeconomic effects, including changes in real gross domestic product (GDP), real private consumption, and the consumer price index, decline by less than 0.05% in all countries for Scenarios 1 and 2. As discussed previously, we expect nominal impacts at the global scale given the relatively small increases in international ballast water treatment costs, and more importantly, because water transport only accounts for relatively minor contributions to total GDP in all countries. For example, water transport contributes 0.2% of GDP in the U.S., and 0.5% of world GDP in total. Stricter BWM regulation applied to all U.S. ports in Scenario 2 negatively affects U.S. economic welfare[5] and negatively affects economic welfare of other countries by a lesser degree; in both cases, the welfare impacts are nominal. Economic welfare in the U.S. decreases by $1.5 billion (-0.01%) in Scenario 1 and $15 billion (-0.11%) in Scenario 2 (Table 4). Welfare losses are less than 0.2% for all countries considered, and Australia, Canada, and Mexico experience higher economic welfare reductions in Scenario 2 resulting from higher transport costs.

[5] We measure economic welfare in terms of equivalent variation, a money-metric measure of economic well-being associated with changes in prices.

Table 4. Changes in economic welfare across countries.

| | Scenario 1 | | Scenario 2 | |
|---|---|---|---|---|
| | % change | Absolute change ($ million) | % change | Absolute change ($ million) |
| Australia | -0.03 | -300 | -0.09 | -1,100 |
| China | -0.03 | -1,600 | -0.02 | -1,100 |
| Japan | -0.01 | -660 | 0.00 | -200 |
| South Korea | -0.03 | -270 | -0.02 | -180 |
| Singapore | 0.00 | -11 | 0.02 | 40 |
| Malaysia | -0.06 | -140 | -0.06 | -140 |
| Taiwan | -0.04 | -170 | -0.03 | -130 |
| USA | -0.01 | -1,500 | -0.11 | -15,000 |
| Canada | -0.04 | -600 | -0.18 | -2,800 |
| Mexico | -0.02 | -250 | -0.11 | -1,100 |
| Colombia | -0.02 | -49 | -0.06 | -170 |
| Panama | -0.06 | -18 | -0.11 | -34 |
| Venezuela | -0.05 | -140 | -0.11 | -310 |
| Belgium | 0.01 | 39 | 0.03 | 150 |
| Germany | 0.00 | -160 | 0.00 | 110 |
| Spain | -0.01 | -77 | 0.00 | 3 |
| France | 0.00 | 37 | 0.01 | 220 |
| UK | -0.03 | -590 | -0.01 | -200 |
| Netherlands | -0.01 | -84 | 0.00 | -21 |
| South Africa | -0.01 | -43 | -0.01 | -26 |

Note: Economic welfare is measured by equivalent variation. Scenario 1 corresponds to the Consistent IMO BWM Regulation and Scenario 2 represents the U.S. Stricter BWM Regulation. Percentages rounded to two decimals, and dollars rounded to millions, for reader comparison; smaller numbers do not imply precision.
Source: Authors' simulations.

**3.3 Changes in shipping traffic**

In general, changes in shipping traffic resulting from ballast water regulation are small, and the shipping pattern is robust to ballast water regulation. Table 5 reports the routes that have at least 10 shipping voyages changed for one of the ship types. For container vessels, the shipping voyages decline by 21 voyages (of 3640 voyages) at most under both policy scenarios, and most of the percentage changes in voyages are less than 1%. The percentage changes in container vessel traffic from South Korea to China and Canada to the U.S. are the same (0.2%), but the absolute voyage change from South Korea to China is much higher (8133 voyages). This is due to the high current voyage numbers from South Korea to China. Stricter U.S. BWM regulation,

Scenario 2, has the biggest impact, from a percentage change perspective, on the shipping routes for container vessels from Canada and Venezuela to the U.S.

The changes in bulk vessel traffic are almost the same under both scenarios between Asian countries, while bulk vessel traffic decreases more in Scenario 2 for North American routes, such as Canada to the U.S. (voyages decrease by 5 to 91 trips), Mexico to the U.S. (voyages decrease by 4 to 52 trips), and the U.S. to Canada (voyages decrease by 20 to 45 trips).

Some routes experience larger decreases for tanker traffic under both policy scenarios, such as Malaysia to Singapore, Singapore to Malaysia, and Netherlands to the UK. Scenario 2 brings a much larger traffic change than Scenario 1 for the routes from Mexico to the U.S., (voyages decrease by 4 to 40 trip) and the U.S. to Mexico (voyages decrease by 1 to 16 trips). The stricter ballast water treatment standards in the U.S. under Scenario 2 will reduce the species invasion risk to the U.S., and decreased shipping traffic will further decrease species introduction risk by limiting the number of voyages. However, one exception to this is for voyages from the U.S. to Columbia, where the decline in tanker traffic is smaller when the U.S. adopts stricter regulation in Scenario 2.

Table 5. Country pairs with largest changes in shipping traffic

| Country pair | Container vessel | | | | | Bulk vessel | | | | | Tanker | | | | |
|---|---|---|---|---|---|---|---|---|---|---|---|---|---|---|---|
| | Current Voyage number | Scenario 1 | | Scenario 2 | | Current Voyage number | Scenario 1 | | Scenario 2 | | Current Voyage number | Scenario 1 | | Scenario 2 | |
| | | Number reduced | % | Number reduced | % | | Number reduced | % | Number reduced | % | | Number reduced | % | Number reduced | % |
| KOR/CHN | 8,123 | 14 | 0.2 | 15 | 0.2 | 1,368 | 11 | 0.8 | 11 | 0.8 | 3,259 | 13 | 0.4 | 13 | 0.4 |
| JPN/CHN | 5,206 | 14 | 0.3 | 17 | 0.3 | 872 | 3 | 0.3 | 3 | 0.3 | 1,300 | 2 | 0.2 | 2 | 0.2 |
| TWN/CHN | 6,366 | 10 | 0.2 | 11 | 0.2 | 903 | 9 | 1.0 | 9 | 1.0 | 1,230 | 6 | 0.5 | 6 | 0.5 |
| CHN/JPN | 6,077 | 19 | 0.3 | 19 | 0.3 | 926 | 8 | 0.9 | 10 | 1.1 | 1,200 | 12 | 1.0 | 10 | 0.8 |
| KOR/JPN | 4,456 | 15 | 0.3 | 16 | 0.4 | 1,031 | 16 | 1.5 | 17 | 1.6 | 3,671 | 12 | 0.3 | 10 | 0.2 |
| SGP/MYS | 3,640 | 21 | 0.6 | 22 | 0.6 | 533 | 8 | 1.5 | 8 | 1.5 | 5,729 | 29 | 0.5 | 26 | 0.5 |
| CAN/USA | 625 | 1 | 0.2 | 10 | 1.6 | 1,535 | 5 | 0.3 | 91 | 5.9 | 778 | 0 | 0 | 5 | 0.6 |
| CHN/TWN | 4,664 | 8 | 0.2 | 7 | 0.2 | 956 | 8 | 0.8 | 10 | 1.1 | 1,374 | 17 | 1.2 | 17 | 1.2 |
| CHN/KOR | 7,537 | 3 | 0 | 3 | 0 | 2,102 | 18 | 0.8 | 22 | 1.1 | 3,797 | 14 | 0.4 | 11 | 0.3 |
| USA/CAN | 801 | 2 | 0.2 | 5 | 0.6 | 1,565 | 20 | 1.3 | 45 | 2.9 | 840 | 0 | 0 | 0 | 0 |
| CHN/AUS | 292 | 0 | 0 | 1 | 0.3 | 3,747 | 6 | 0.2 | 24 | 0.6 | 40 | 0 | 0 | 0 | 0 |
| JPN/AUS | - | - | - | - | - | 1,758 | 3 | 0.2 | 12 | 0.7 | 77 | 0 | 0 | 0 | 0 |
| MEX/USA | 700 | 0 | 0 | 1 | 0.1 | 757 | 4 | 0.5 | 52 | 6.9 | 1,738 | 4 | 0.2 | 40 | 2.3 |
| MYS/SGP | 5,000 | 7 | 0.1 | 7 | 0.1 | 570 | 4 | 0.7 | 4 | 0.7 | 5,946 | 50 | 0.8 | 51 | 0.9 |
| NLD/GBR | 3,287 | 4 | 0.1 | 4 | 0.1 | 988 | 0 | 0 | 1 | 0.1 | 5,103 | 36 | 0.7 | 43 | 0.8 |
| USA/COL | 271 | 1 | 0.4 | 2 | 0.7 | 122 | 0 | 0 | 1 | 0.8 | 438 | 19 | 4.3 | 3 | 0.7 |
| USA/MEX | 381 | 1 | 0.3 | 3 | 0.8 | 612 | 2 | 0.3 | 5 | 0.8 | 1,773 | 1 | 0.1 | 16 | 0.9 |

Note: Scenario 1 corresponds to the Consistent IMO BWM Regulation and Scenario 2 represents the U.S. Stricter BWM Regulation. The routes shown have at least 10 shipping voyages in traffic change for one of the ship types. No container vessel traffic exists from Japan to Australia. Percentages rounded to one decimal. Countries are shown in International Organization for Standardization (ISO) codes: https://unstats.un.org/unsd/tradekb/knowledgebase/country-code.
Source: Authors' simulations

### 3.4 Comparison with studies on environmental policies

The environmental policies for BWM examined in this study target increased costs for marine transport. Spillover effects are then spread throughout the economy. Since the marine transport sector only accounts for small contributions to individual country GDP, and comprise 0.5% of world GDP, BWM policies we investigate have minor economic impacts at aggregate levels for trade, output and national economic welfare for each country in this study. Other environmental or climate-related policies, such as carbon taxes, emissions trading schemes (ETS), and energy taxes, are found to impose higher costs on national economies than the costs we estimate for BWM scenarios. For example, Adams et al. estimated that an international ETS to decrease global emission levels by 5% in 2050 relative to 2000 levels would decrease real GDP of Australia by 1.1% by 2030 relative to the base case scenario, while real household consumption and imports would decline by 1.5% and 2%, respectively (Adams et al., 2014). Using different assumptions and emission quotas of an ETS to achieve the 2030 target of 28% below the 2005 level in Australia, Nong et al. found a decline of 1.6% of Australian real GDP by 2030 relative to the baseline when examining the impacts of such ETS in Australia. Real exports and imports for Australia would also fall by 2.77% and 2.94%, respectively (Nong et al., 2017). Carbon tax studies in South Africa find real GDP decreases by 1-1.59% under different carbon tax rates from $9.15 to $30 per ton of carbon dioxide equivalent ($CO_2$e) (Alton et al., 2014, Nong, 2020). An international carbon tax study of $50 per ton of $CO_2$e also indicates that real GDP would decline by 0.94–1.28% for the United States, 0.66–1.54% for China, 0.34–0.57% for India, 0.44–0.55% for Brazil, and 0.12% for the European Union (Nong and Simshauser, 2020).

Energy tax scenarios impose much larger economic costs than explored in our analysis and find larger unfavorable impacts. Tax rates of 5–15% on energy, for example, simulate a decrease of 0.27–1.13% in Chinese GDP (Peng et al., 2019). Energy taxes associated with renewable energy development and proper revenue recycling mechanisms equal to a 7% tax rate would decrease real GDP of Spain by 0.08–0.1% (Freire-González and Puig-Ventosa, 2019). Nong et al. also found that substantial increases in energy taxes of 50% for coal and 33% for petroleum products would reduce real GDP of Vietnam by 1.05% and 2.23%, respectively, with private consumption declines ranging from 0.34 to 1.06% (Nong et al., 2019).

It is worth noting that the impacts of a carbon tax, ETS, or energy tax vary across countries, depending on several key factors. Economic effects depend on the magnitudes of the costs imposed, the market structures and market shares of energy/emission-intensive sectors and expenditure shares for energy commodities, details of revenue recycling mechanisms, and on the extent of sector coverage in each country and potential policy. These environmental or climate change policies in the literature focus on reduced greenhouse gas emission levels and apply to most (if not all) sectors in the economy, while BWM regulations impose direct costs on the marine transport sector.

Both the costs imposed and economic impacts of BWM regulations are also small relative to other transport costs studies. For example, Bekkers et al. (2018) found that melting Arctic sea ice enabling the Northern Sea Route with transportation cost reductions between Northern Europe and East Asia by 20–30% for all goods transported by sea would affect bilateral real exports from China by -0.51% to 11.98%, and Japan by -0.49% to 16.23%. Real GDP is also affected differently from -0.1% to 0.4% across countries (Bekkers et al., 2018). On the same topic, but also accounting for the additional new route of the Northwest Passage connecting Northeast Asia and the U.S. east coast, Countryman et al. (2016) found major impacts on trade of different countries resulting from decreased shipping costs given shorter shipping distances through the Arctic. They found that transportation costs between China, Japan, South Korea, Taiwan and the U.S. east coast would decrease by 6.8–11.2% for all sea transport, while transport costs for routes between Europe and Northeast Asian countries would decrease even further. For example, shipping costs between East Asian nations and Germany would decrease by 19.9–39.4%. As a result, U.S. exports were simulated to increase by 13.5% to China, 13.5% to Japan, and 5% to South Korea. Real GDP changes across countries vary from -0.5% to 1%, and are notably higher than the GDP effects of BWM because the costs changes are of larger magnitudes and affect all goods transported by sea between these regions (Countryman et al., 2016).

This policy review shows that the costs imposed by the aforementioned policies are relatively much higher than those derived from examining BWM policies in the present study. Accordingly, results from other environmental policy studies are substantially higher at aggregate and sectoral levels than the economic effects of alternate BWM policy scenarios investigated in this work. However, it is important to note that while the costs and resulting economic impacts of BWM are minor across countries at aggregate levels, potential environmental benefits from BWM may be more widespread across countries and sectors and merit investigation in future work. Further research to explore the benefits of improved environmental quality relative to the costs that we estimate is warranted.

## 4. Conclusion and implications

This study provides a noteworthy contribution to explore the viability to drive ballast water management policies forward, as well as to integrated economic modeling research. We investigate the impacts of international ballast water policy on transport costs, international trade, national economies and global shipping patterns. Our findings are consistent with fundamental trade theory and show that the costs of environmental action through BWM in the shipping industry have small economic consequences at aggregate levels but vary in magnitude for bilateral trade of specific commodities between certain countries.

Specifically, our work provides three key findings: 1) current BWM regulation that is uniform across countries (Scenario 1) has nominal negative economic impacts across countries; 2) the proposed scenario of stricter BWM regulation imposed for transport to/from U.S. ports (Scenario 2) has minor effects on trade and national economies at aggregate levels, but causes noteworthy trade reductions of certain commodities between some countries; 3) BWM policies under uniform and stricter standards lead to modest changes in global shipping patterns.

Findings suggest that international ballast water treatment regulation, even when the U.S. applies stricter standards, has minor effects on the global economy when considering changes in international trade, economic welfare, and shipping patterns, yet increased environmental quality can be achieved. However, it is important to consider both aggregate and bilateral trade effects when imposing increased transport costs to achieve invasive species risk reduction through BWM since considerable sector, and country-specific effects are masked when assessing economic effects at aggregate levels across countries.

Results show that changes in shipping costs vary across vessel types, as well as between trade partners because of different trade volumes and values. We find that shipping costs between the U.S. and its trade partners only change slightly in Scenario 1 and are higher for some routes in Scenario 2 due to stricter regulation at U.S. ports. Such cost changes yield relatively small adverse impacts on the global economy for both scenarios due to small shares of the marine transportation sector in the overall economy in each country. However, the trade impacts for certain commodities between countries shipped by different vessel types are highly affected due to relatively high increased shipping costs of vessels for certain shipping routes, including trade between Australia and the U.S., for example. Due to modest changes in total trade values/volumes, there are also only minor changes in shipping voyages at the global scale.

Both uniform and stricter BWM standards negatively affect global economic welfare by small rates, driven by the small share of total global GDP (0.5%) attributed to the marine transport sector. Stricter regulation increases the transportation costs for commodities originating from any trade partner to the U.S. market. However, the U.S. is a large trade partner of many countries around the world, and the costs of BWM are shared across voyages for each vessel type. As a result, the overall marine transport services costs in all countries increase at higher rates in Scenario 2 relative to the increases in Scenario 1 – when all countries apply an international BWM standard regulation. Since costs of commodities exported to the U.S. market increase, trade is diverted as exporters find other countries for their export products. One key example of trade diversion can be seen for Australian producers of forest products and foods, textiles, metal and chemicals, and machine and equipment, which increased exports to all other countries at the expense of the U.S. in Scenario 2 when shipping to the U.S. becomes substantially more expensive for this route under stricter BWM regulation. However, trade effects from these cost signals may result in perturbations to the volume of trade (as per GTAP output) or to other shifts in shipping patterns or fleet operations not modeled here

Economic welfare losses are small for all countries when compared to the size of each national economy. U.S. economic welfare losses are the highest, though are nominal, totaling -$1,500 in Scenario 1 and -$14,800 in Scenario 2. Other key exporters to the U.S. including Australia, Canada, and Mexico face decreased economic welfare by lesser magnitudes than the U.S. Results confirm that stricter BWM standards applied to U.S. ports affect U.S. producers the most, followed by key U.S. trading partners. While our work focuses on the economic costs of BWM regulation, research to investigate the benefits of increased environmental quality due to stricter BWM standards is warranted for further study to provide decision makers with an evaluation of the net benefits of BWM and the cost-effectiveness thresholds of potential policy measures.

## Appendix A: Map of commodity types with vessel types

The GTAP model includes 57 commodity types that we aggregate to 17 sector groups. The 17 sectors are mapped to 8 vessel types according to industry practice (UNCTAD, 2017).

Table A1. GTAP commodity mapping to vessel types

| GTAP Commodities | Aggregated commodities | Vessel types | Group number |
|---|---|---|---|
| OCR (Crops nec) | Crop products | Bulk | 1 |
| WHT (Wheat) | Wheat & grains | Bulk | 2 |
| GRO (Cereal grains nec) | Wheat & grains | Bulk | 2 |
| SGR (Sugar) | Forestry and foods | Con/Bulk | 3 |
| PDR (Paddy rice) | Forestry and foods | Container | 3 |
| OSD (Oil seeds) | Forestry and foods | Container | 3 |
| C_B (Sugar cane, sugar beet) | Forestry and foods | Container | 3 |
| PFB (Plant-based fibers) | Forestry and foods | Container | 3 |
| WOL (Wool, silk-worm cocoons) | Forestry and foods | Container | 3 |
| FRS (Forestry) | Forestry and foods | Container | 3 |
| PCR (Processed rice) | Forestry and foods | Container | 3 |
| OFD (Food products nec) | Forestry and foods | Container | 3 |
| B_T (Beverages and tobacco products) | Forestry and foods | Container | 3 |
| CTL (cattle, sheep and goats, horses) | Life animal | Livestock carrier | 4 |
| V_F (Vegetables, fruit, nuts) | Meats & vegetable | Refrigeration | 5 |
| FSH (Fishing) | Meats & vegetable | Refrigeration | 5 |
| CMT (Bovine meat products) | Meats & vegetable | Refrigeration | 5 |
| OMT (Meat products nec) | Meats & vegetable | Refrigeration | 5 |
| VOL (Vegetable oils and fats) | Meats & vegetable | Refrigeration | 5 |
| MIL (Dairy products) | Meats & vegetable | Refrigeration | 5 |
| OAP (Animal products nec) | Meats & vegetable | Refrigeration | 5 |
| RMK (Raw milk) | Meats & vegetable | Refrigeration | 5 |
| TEX (Textiles) | Textiles | Container | 6 |
| WAP (Wearing apparel) | Textiles | Container | 6 |
| LEA (Leather products) | Textiles | Container | 6 |
| LUM (Wood products) | Textiles | con/bulk | 6 |
| PPP (Paper products, publishing) | Textiles | Container | 6 |
| CRP (Chemical, rubber, plastic) | Metal & chemicals | con/bulk | 7 |
| OMN (Minerals nec) | Metal & chemicals | con/bulk | 7 |
| NMM (Mineral products nec) | Metal & chemicals | con/bulk | 7 |
| I_S (Ferrous metals) | Metal & chemicals | con/bulk | 7 |
| NFM (Metals nec) | Metal & chemicals | con/bulk | 7 |
| FMP (Metal products) | Metal & chemicals | con/bulk | 7 |

Table A1. GTAP commodity mapping to vessel types (continued)

| | | | |
|---|---|---|---|
| ELE (Electronic equipment) | Machine & equipment | Container | 8 |
| OME (Machinery and equipment nec) | Machine & equipment | Container | 8 |
| OMF (Manufactures nec) | Machine & equipment | Container | 8 |
| OTN (Transport equipment nec) | Machine & equipment | General cargo | 8 |
| MVH (Motor vehicles and parts) | Motor vehicles | Vehicle | 9 |
| ELY (Electricity) | Electricity | | 10 |
| COA (Coal) | Coal | Bulk | 11 |
| P_C (Petroleum, coal products) | Petroleum products | Tanker | 12 |
| OIL (Oil) | Crude oil | Tanker | 13 |
| GAS (Gas) | Natural gas | LG/pipeline | 14 |
| WTP (Water transport) | Water transport | | 15 |
| OTP (Transport nec) | Other transport | | 16 |
| ATP (Air transport) | Other transport | | 16 |
| GDT (Gas manufacture, distribution) | Other Services | | 17 |
| WTR (Water) | Other Services | | 17 |
| CNS (Construction) | Other Services | | 17 |
| TRD (Trade) | Other Services | | 17 |
| CMN (Communication) | Other Services | | 17 |
| OFI (Financial services nec) | Other Services | | 17 |
| ISR (Insurance) | Other Services | | 17 |
| OBS (Business services nec) | Other Services | | 17 |
| ROS (Recreational and other services) | Other Services | | 17 |
| OSG (Public Administration, etc.) | Other Services | | 17 |
| DWE (Dwellings) | Other Services | | 17 |

Source: Commodity cargo and vessels matched and summarized by authors.